\documentclass[a4paper]{jpconf}

\pdfoutput=1

\usepackage{ifpdf}

\usepackage{graphicx}

\usepackage{latexsym}
\usepackage{amsfonts}
\usepackage{amssymb}
\usepackage{amsbsy}
\usepackage{amsmath}   %

\usepackage[usenames,dvipsnames]{pstricks}
\ifpdf
 \else
  \usepackage{epsfig}
  \usepackage{pst-solides3d}
\fi

\def\p{\partial}
\def\dfrac#1#2{{\displaystyle\frac{#1}{#2}}}
\def\stTD#1#2{\hbox to 0em{\mathsurround=0em $\stackrel{#1}{\makebox[0pt]{} #2}$\hss} \phantom{#2}}\def\stscript#1#2{\hbox to 0em{\mathsurround=0em ${\scriptstyle\stackrel{#1}{\makebox[0pt]{} #2}}$\hss} \phantom{#2}}\def\stscriptscript#1#2{\hbox to 0em{\mathsurround=0em ${\scriptscriptstyle\stackrel{#1}{\makebox[0pt]{} #2}}$\hss} \phantom{#2}}
\def\comb#1#2#3{{\mathsurround 0pt\hbox to 0pt {\hspace*{#3}\raisebox{#2}{${#1}$}\hss}}}
\def\combs#1#2#3{{\mathsurround 0pt\hbox to 0pt {\hspace*{#3}\raisebox{#2}{${\scriptstyle #1}$}\hss}}}
\def\combss#1#2#3{{\mathsurround 0pt\hbox to 0pt {\hspace*{#3}\raisebox{#2}{${\scriptscriptstyle #1}$}\hss}}}

\def\df{\mathrm{d}}

\def\p{\partial}
\def\dfrac#1#2{{\displaystyle\frac{#1}{#2}}}
\def\stTD#1#2{\hbox to 0em{\mathsurround=0em $\stackrel{#1}{\makebox[0pt]{} #2}$\hss} \phantom{#2}}\def\stscript#1#2{\hbox to 0em{\mathsurround=0em ${\scriptstyle\stackrel{#1}{\makebox[0pt]{} #2}}$\hss} \phantom{#2}}\def\stscriptscript#1#2{\hbox to 0em{\mathsurround=0em ${\scriptscriptstyle\stackrel{#1}{\makebox[0pt]{} #2}}$\hss} \phantom{#2}}
\def\comb#1#2#3{{\mathsurround 0pt\hbox to 0pt {\hspace*{#3}\raisebox{#2}{${#1}$}\hss}}}
\def\combs#1#2#3{{\mathsurround 0pt\hbox to 0pt {\hspace*{#3}\raisebox{#2}{${\scriptstyle #1}$}\hss}}}
\def\combss#1#2#3{{\mathsurround 0pt\hbox to 0pt {\hspace*{#3}\raisebox{#2}{${\scriptscriptstyle #1}$}\hss}}}

\def\Act{\mathcal{A}}

\def\Vol{\overline{V}}

\def\xxx{\chi}
\def\ffun{\Phi}

\def\dffun{\ffun}

\def\metr{\mathfrak{m}}

\def\metrp{\mathchoice{\comb{-}{-0.9ex}{0ex}\mathfrak{m}}{\comb{-}{-0.9ex}{0ex}\mathfrak{m}}{\combs{-}{-0.75ex}{-0.1ex}\mathfrak{m}}{}{}}

\def\eqdef{\doteqdot}

\def\metrEff{\mathchoice{\combs{\sim}{1ex}{0.2ex}\mathfrak{m}}{\combs{\sim}{1ex}{0.2ex}\mathfrak{m}}{\combss{\sim}{0.66ex}{0.05ex}\mathfrak{m}}{}{}}
\def\metrEffm1{\check{\metrEff}}

\def\df{\mathrm{d}}

\def\metr{\mathfrak{m}}

\def\metrp{\mathchoice{\comb{-}{-0.9ex}{0ex}\mathfrak{m}}{\comb{-}{-0.9ex}{0ex}\mathfrak{m}}{\combs{-}{-0.75ex}{-0.1ex}\mathfrak{m}}{}{}}

\def\bje{{\mathsurround 0pt\lower.0ex\hbox{${\scriptscriptstyle \mathbf{e}}$}\mspace{-3.4mu}\mathbf{j}}}
\def\je{{\mathsurround 0pt\lower.0ex\hbox{${\scriptscriptstyle e}$}\mspace{-4.5mu}j}}
\def\bjm{{\mathsurround 0pt\lower.0ex\hbox{${\scriptscriptstyle \mathbf{m}}$}\mspace{-5.6mu}\mathbf{j}}}
\def\jm{{\mathsurround 0pt\lower.0ex\hbox{${\scriptscriptstyle m}$}\mspace{-7.0mu}j}}

\def\p{\partial}

\def\eqdef{\doteqdot}

\def\OOO#1#2{\mathcal{O}\!\left(#1\right)_{#2}}

\def\Act{\mathcal{A}}

\def\Vol{\overline{V}}

\def\xxx{\chi}

\def\ffun{\Phi}

\def\dffun{\ffun}

\def\OOO#1#2{\mathcal{O}\!\left(#1\right)_{#2}}

\begin{document}

\title[About toroidal soliton-particle of extremal space-time film]{About toroidal soliton-particle\\ of extremal space-time film}
\author{Alexander A. Chernitskii}

\address{$^1$ Department  of Mathematics\\ St. Petersburg State Chemical Pharmaceutical University\\Prof. Popov str. 14, St. Petersburg, 197022, Russia}
\address{$^2$ A. Friedmann Laboratory for Theoretical Physics\\St. Petersburg, Russia}

\ead{AAChernitskii@mail.ru}

\begin{abstract}
Nonlinear field model of extremal space-time film is considered.
Its space-localized solution in toroidal coordinates with periodic dependence in time is investigated.
A field configuration having a form of the twisted lightlike soliton moving along the ring is considered.
An iterative algorithm for obtaining the solution in the form of formal power series in one toroidal variable is proposed.
It is discovered that the time dependence in the toroidal solution has a deeply embedded character.
Thus the full energy and angular momentum of the solution converge at space infinity.
\end{abstract}

\section{Introduction}
\label{introd}
Let us start from the concept of unified field theory which is known for a long time.
It is well known also that the creator of relativity theory A. Einstein was a proponent of this concept \cite{Einstein1953aE}.
Also many famous researchers worked in this direction.

According to this concept
the elementary particles must be represented as space-localized solutions of the model nonlinear equations.
A relatively weak progress in this direction of investigation is connected with the appropriate extraordinary mathematical difficulties.

Recent defined advancement in the part of finding of exact solutions for a nonlinear field model \cite{Chernitskii2018a} can stimulate a further progress
in the representation of elementary particles by soliton solutions of field equations.

A nonlinear space-time scalar field model considered here is known for a long time sufficiently.
This model is related to well known Born--Infeld nonlinear electrodynamics \cite{Chernitskii2004a,Chernitskii2012be},
and it is sometimes called Born--Infeld type scalar field model \cite{BarbChern1967-1e}.

We call this model the extremal space-time film one \cite{Chernitskii2018a}, because it is a relativistic generalization of the minimal surface or
minimal thin film model in three-dimensional space.

According to recent results \cite{Chernitskii2018a,Chernitskii2016a} we can consider a correspondence between the first-order twisted lightlike solitons of extremal space-time film and photons.
In particular,
the equilibrium energy spectral density for ideal gas of these solitons
has the form of Planck distribution in some approximation.

A correspondence between higher-order twisted lightlike solitons and real elementary particles will be investigated later. In particular,
we keep in mind the possible correspondence between these solitons and neutrinos.

Now we continue the investigation of time-periodic solitons with the general toroidal symmetry as massive elementary particles.
Such field configurations in toroidal coordinates was considered in the framework of Born--Infeld nonlinear
electrodynamics \cite{Chernitskii200814}.%

In the present work we consider the problem formulation and some new results concerning the representation of leptons by ringed solitons of extremal space-time film.

\section{Extremal space-time film}
\label{exatf}

Let us consider the extremal variational principle with the action having the world volume form:
 \begin{equation}
\label{35135655}
\Act  =\int_{\Vol}\!\sqrt{|\mathfrak{M}|}\;(\mathrm{d}x)^{4}
\;,\quad
\mathfrak{M} \eqdef \det(\mathfrak{M}_{\mu\nu})
\;,\quad
\mathfrak{M}_{\mu\nu} = \metr_{\mu\nu} + \xxx^2\,\frac{\p \ffun}{\p x^{\mu}}\,\frac{\p \ffun}{\p x^{\nu}}
\;,
\end{equation}
where $\left(\df x\right)^{4} \eqdef \df x^{0}\df x^{1}\df x^{2}\df x^{3}$,
 $\Vol$ is a space-time volume,
 $\metr_{\mu\nu}$ are components of metric tensor for flat four-dimensional space-time,
 $\ffun$ is scalar real field function,
 $\xxx$ is dimensional constant.
 The Greek indices take values $\{0,1,2,3\}$.
The tensor $\mathfrak{M}_{\mu\nu}$  can be called also the world tensor.

The model (\ref{35135655}) can be considered as a relativistic generalization of
the mathematical model of two-dimensional minimal thin film
in the tree-dimensional space of our everyday experience.

The field equation in Cartesian coordinates has the following form:
 \begin{equation}
 \label{371394071}
 \left(\metrp^{\mu\nu}\left(1 + \xxx^{2}\,\metrp_{\sigma\rho}\,\ffun^{\sigma}\,\ffun^{\rho}\right) - \xxx^{2}\,\dffun^{\mu}\,\dffun^{\nu}\right)
 \frac{\p^{2}\,\ffun}{\p x^{\mu}\,\p x^{\nu}}  = 0
 \;,\quad
\ffun^{\alpha}  \eqdef \metrp^{\alpha\beta}\,\frac{\p \ffun}{\p x^{\beta}}
\;,
 \end{equation}
where $\metrp^{\mu\nu}$ is Minkowski metric with signature $\{+,-,-,-\}$ or $\{-,+,+,+\}$.

\section{Space-time film  equation in toroidal coordinates}
\label{ectc}

\begin{figure}[h]
\begin{minipage}{17.7pc}
{\unitlength 1.0mm
\begin{picture}(50,63)
\put(-12,0){
\ifpdf
 \put(13,0){\includegraphics[width=1.1\linewidth]{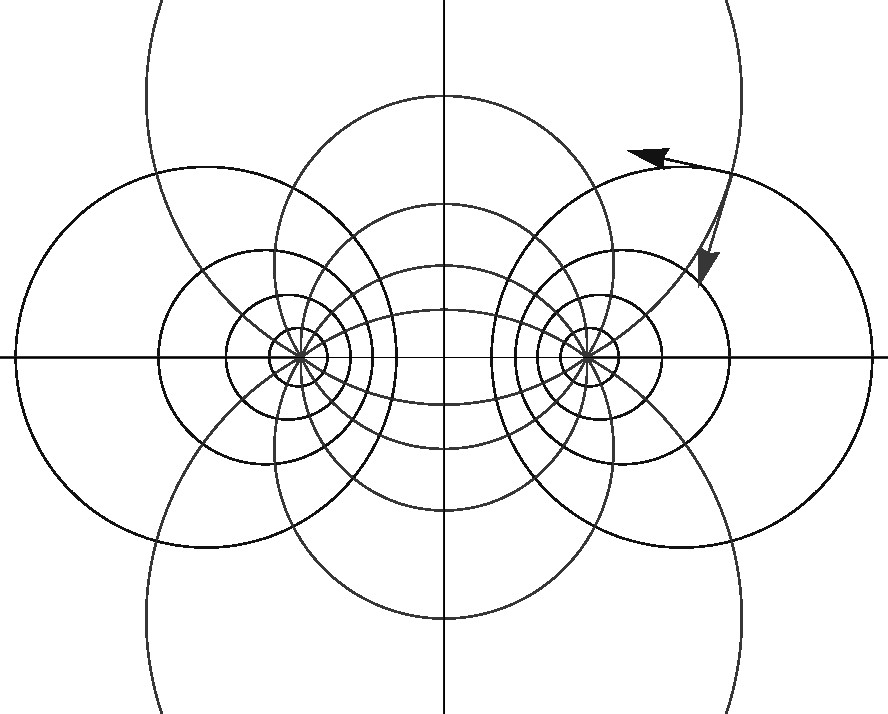}}
\else
 \put(13,0){\includegraphics[width=1.1\linewidth]{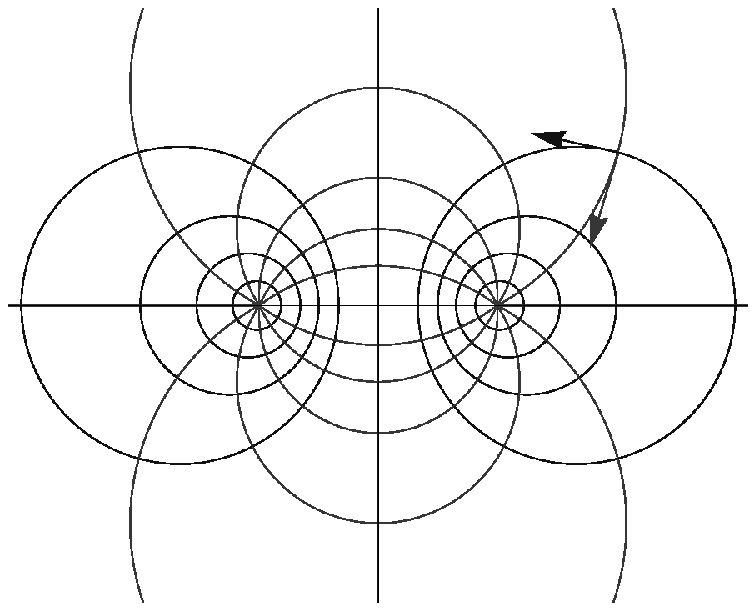}}
\fi
\put(-2.35,2.5){
\put(56,61){\oval(10,10)[b]}
\put(61,62){\vector(0,1){0.1}}
\put(62,62){\makebox(0,0)[lc]{$\varphi$}}
}
\put(81.5,41){\makebox(0,0)[cc]{$\mathbf{e}_\kappa$}}
\put(75,54){\makebox(0,0)[cc]{$\mathbf{e}_\upsilon$}}
\put(-32,-1.2){
\put(69,8){\makebox(0,0)[lc]{$\kappa=\infty$}}
\put(67,6){\line(1,0){14.8}}
\put(81.8,6){\vector(-1,3){9}}
}
\put(-5.5,-1.2){
\put(69,8){\makebox(0,0)[lc]{$\kappa=\infty$}}
\put(67,6){\line(1,0){14.8}}
\put(81.8,6){\vector(-1,3){9}}
}
\put(51.0,51.5){\makebox(0,0)[lc]{$\kappa = 0$}}
\put(82,35){\makebox(0,0)[lc]{$\upsilon=0$}}
\put(51.0,34.5){\makebox(0,0)[lc]{$\upsilon=\pi$}}
\put(51.0,31){\makebox(0,0)[lc]{$\upsilon=-\pi$}}
}
\end{picture}
}
\caption{\label{labelFig1}Section $\varphi = 0 \cup \varphi = \pi$ for toroidal coordinates.}
\end{minipage}\hspace{2pc}%
\begin{minipage}{17.7pc}
\begin{picture}(150,190)
\ifpdf
 \put(-3,90){\includegraphics[width=10pc]{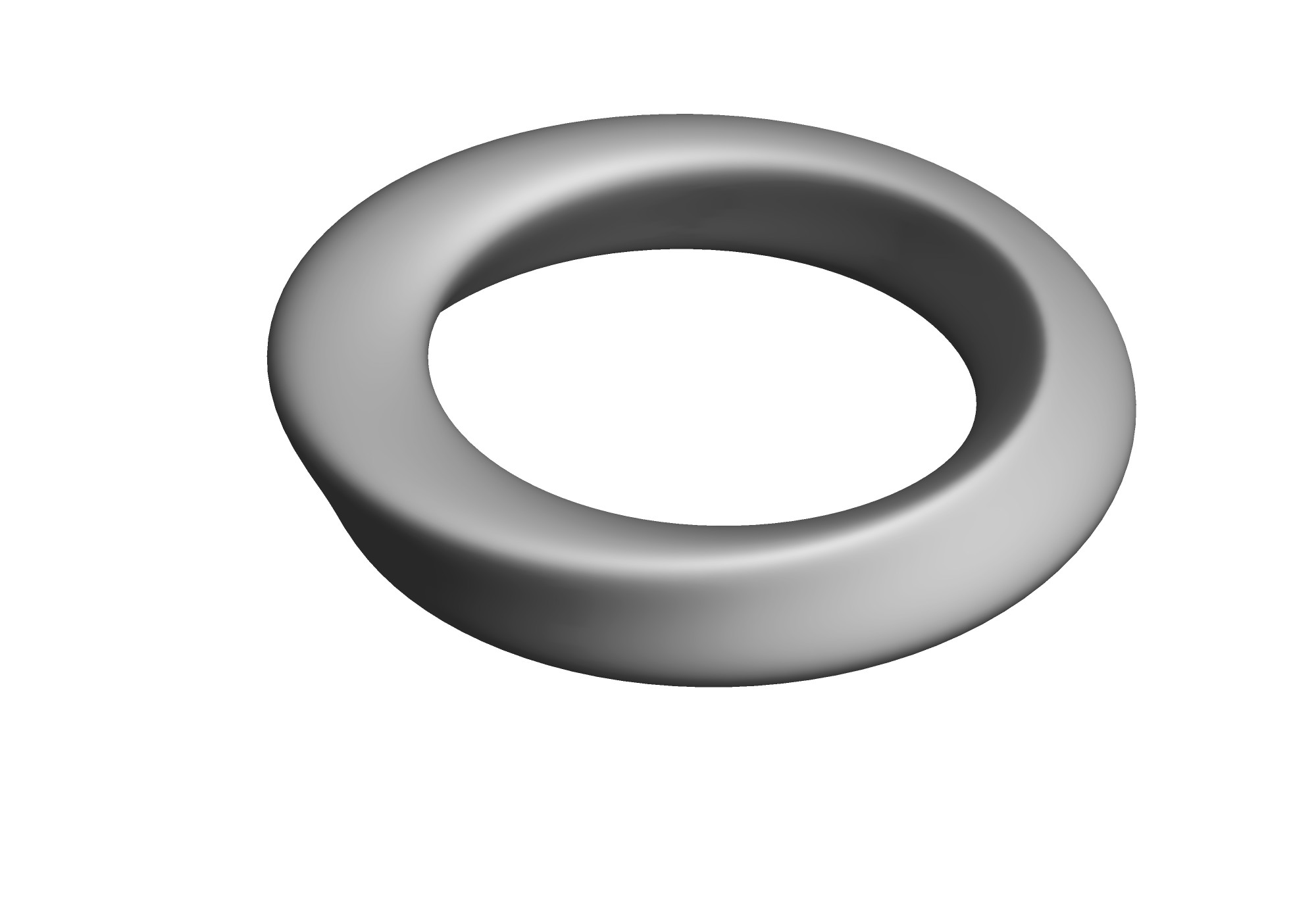}}
 \put(105,90){\includegraphics[width=10pc]{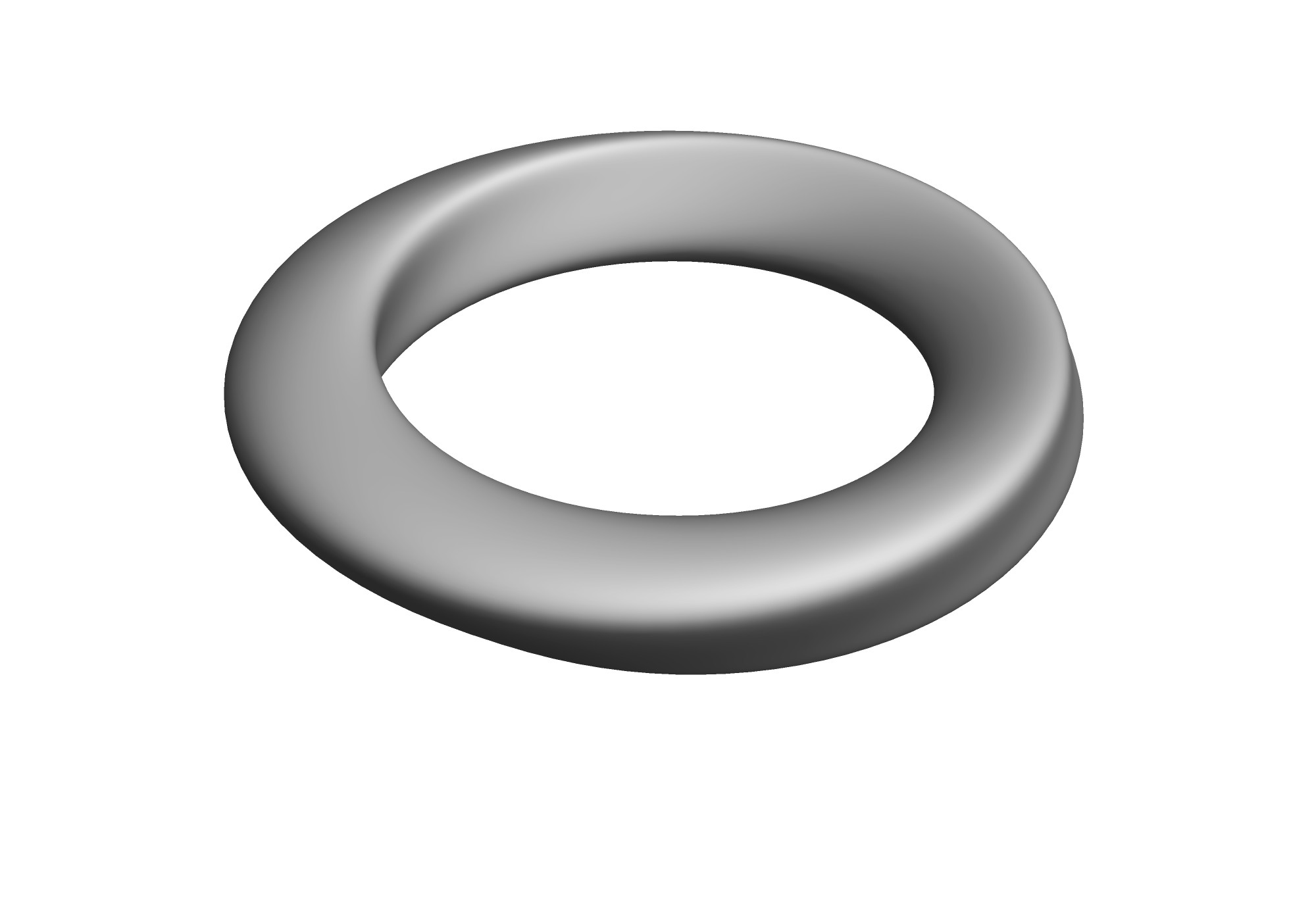}}
 \put(-3,5){\includegraphics[width=10pc]{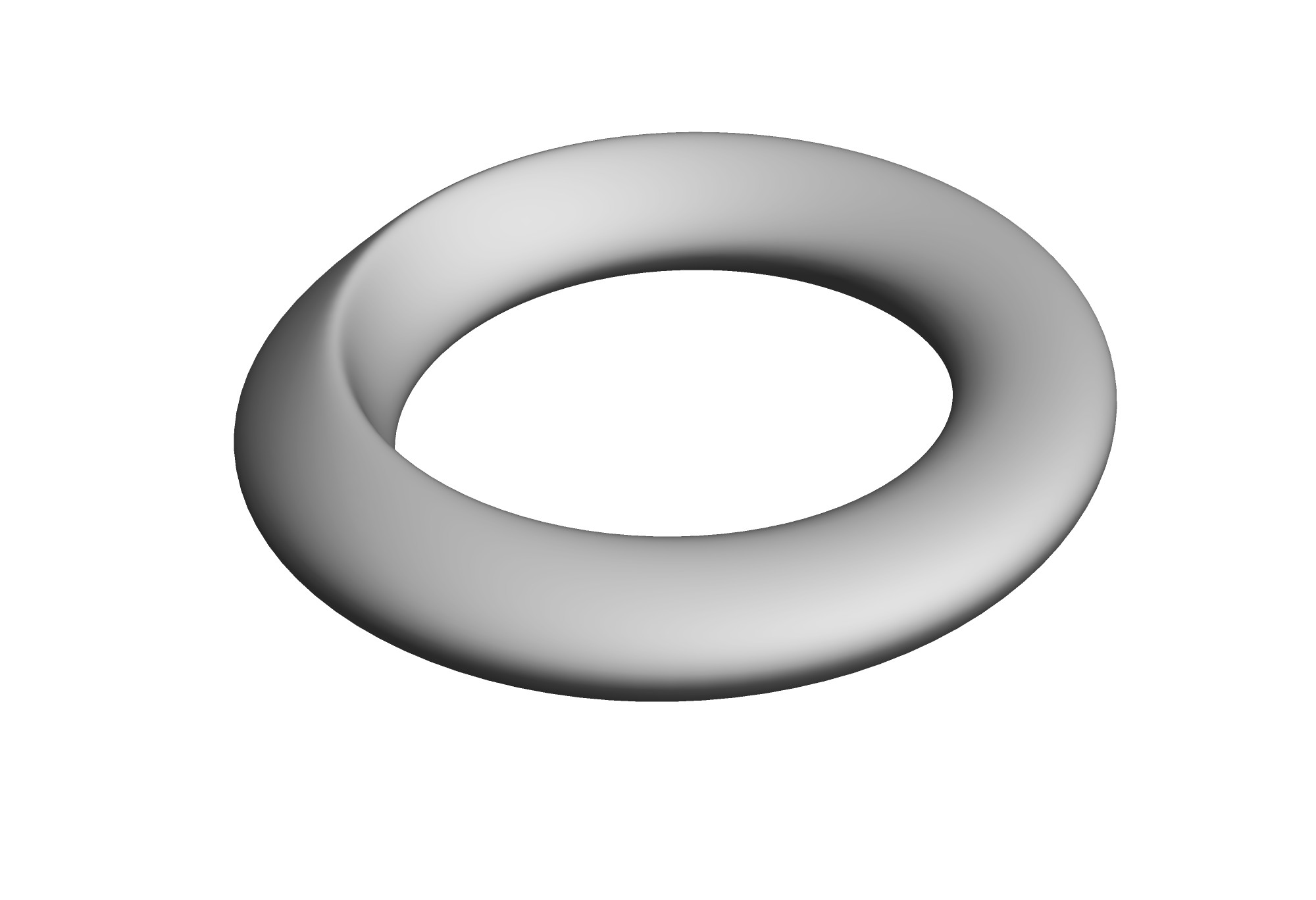}}
 \put(105,5){\includegraphics[width=10pc]{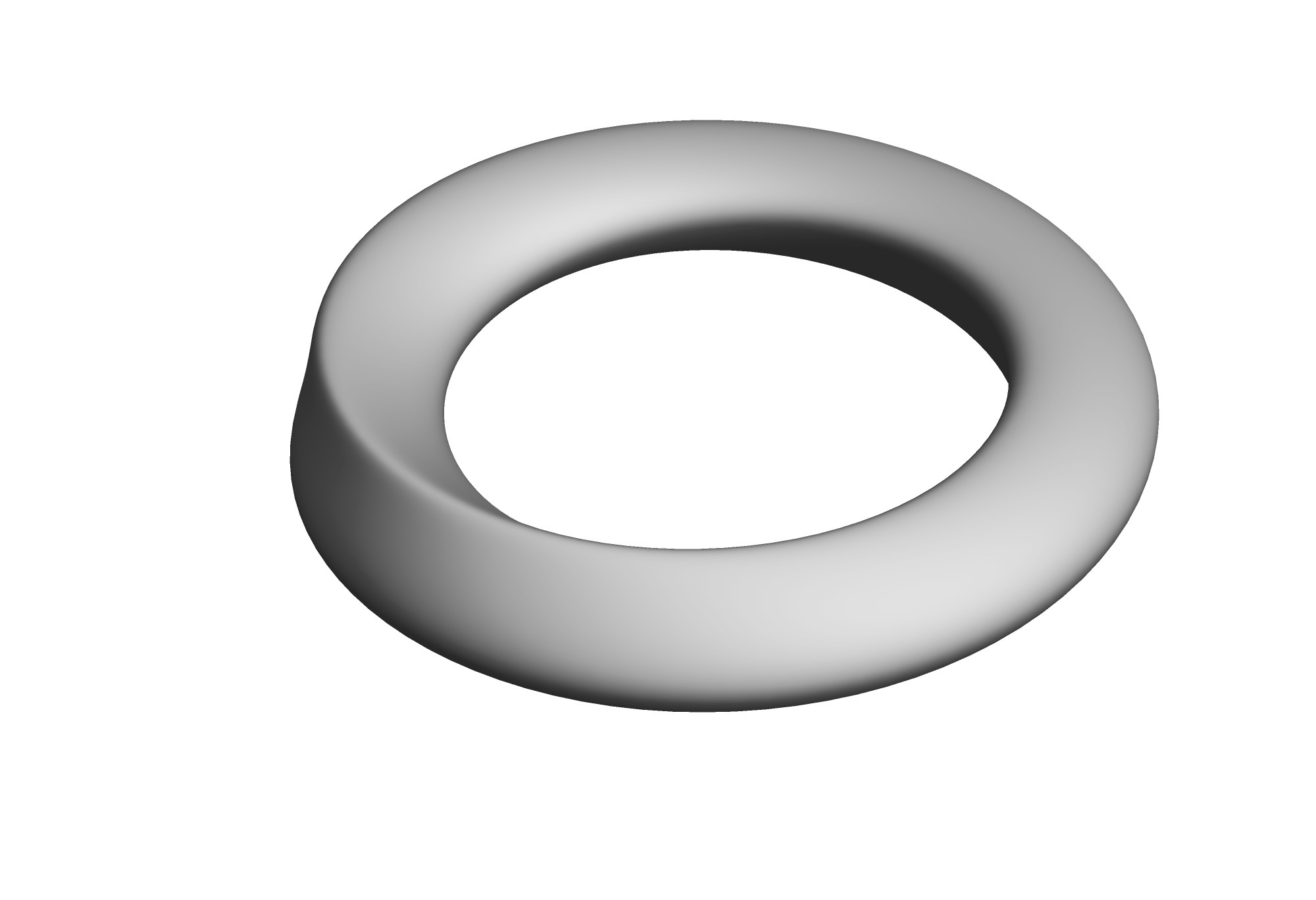}}
\else
 \put(-3,90){\includegraphics[width=10pc]{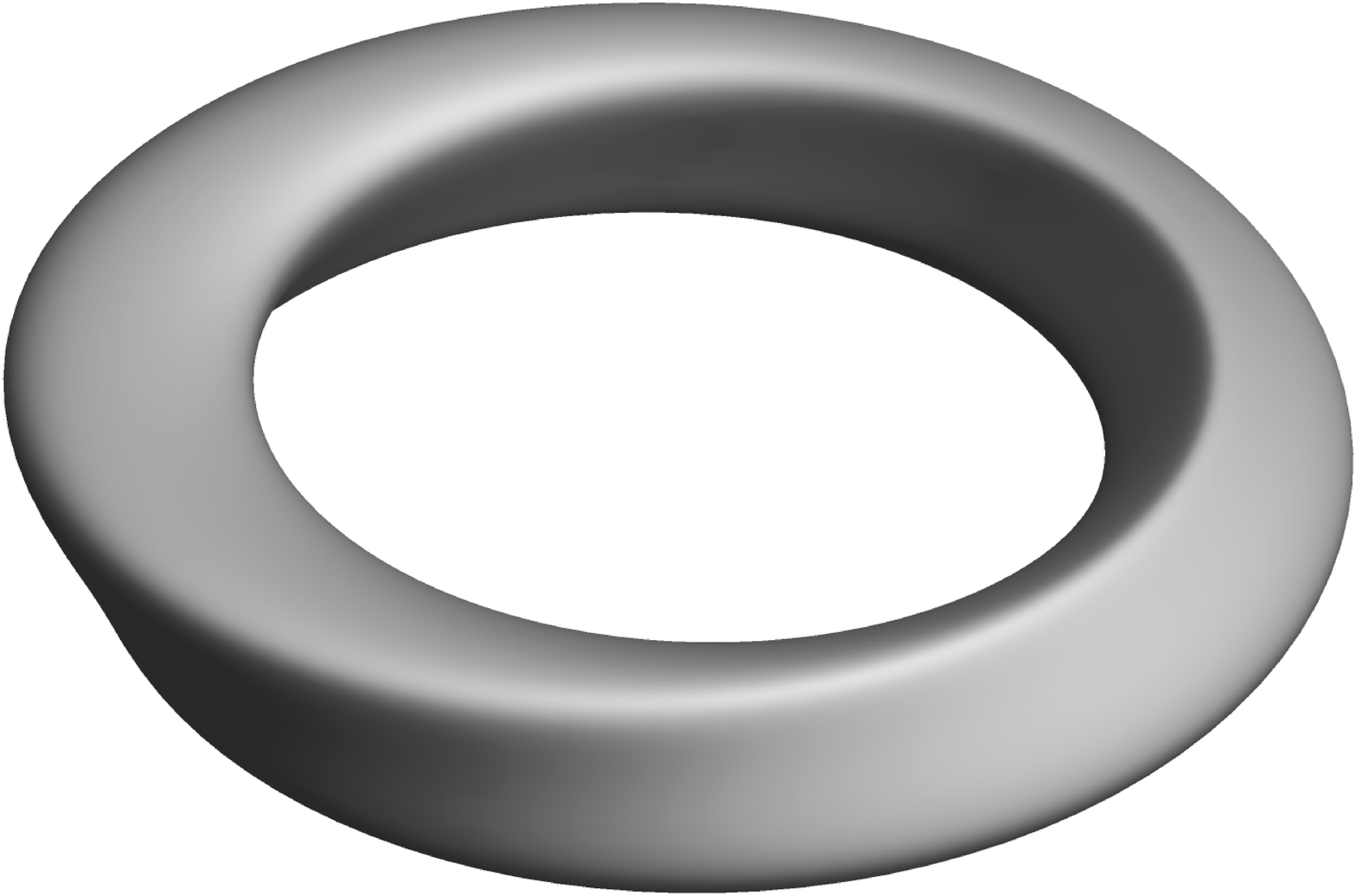}}
 \put(105,90){\includegraphics[width=10pc]{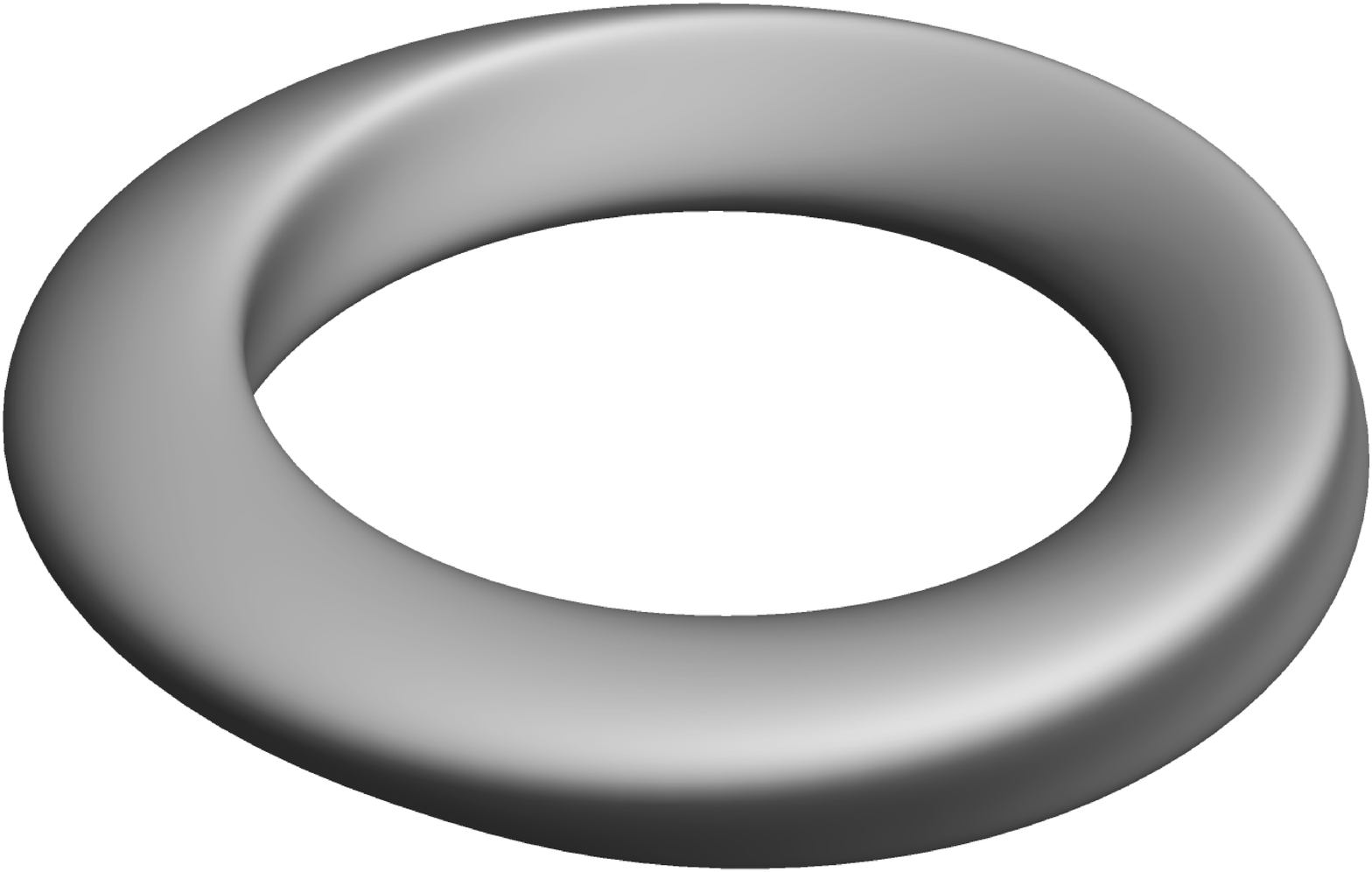}}
 \put(-3,5){\includegraphics[width=10pc]{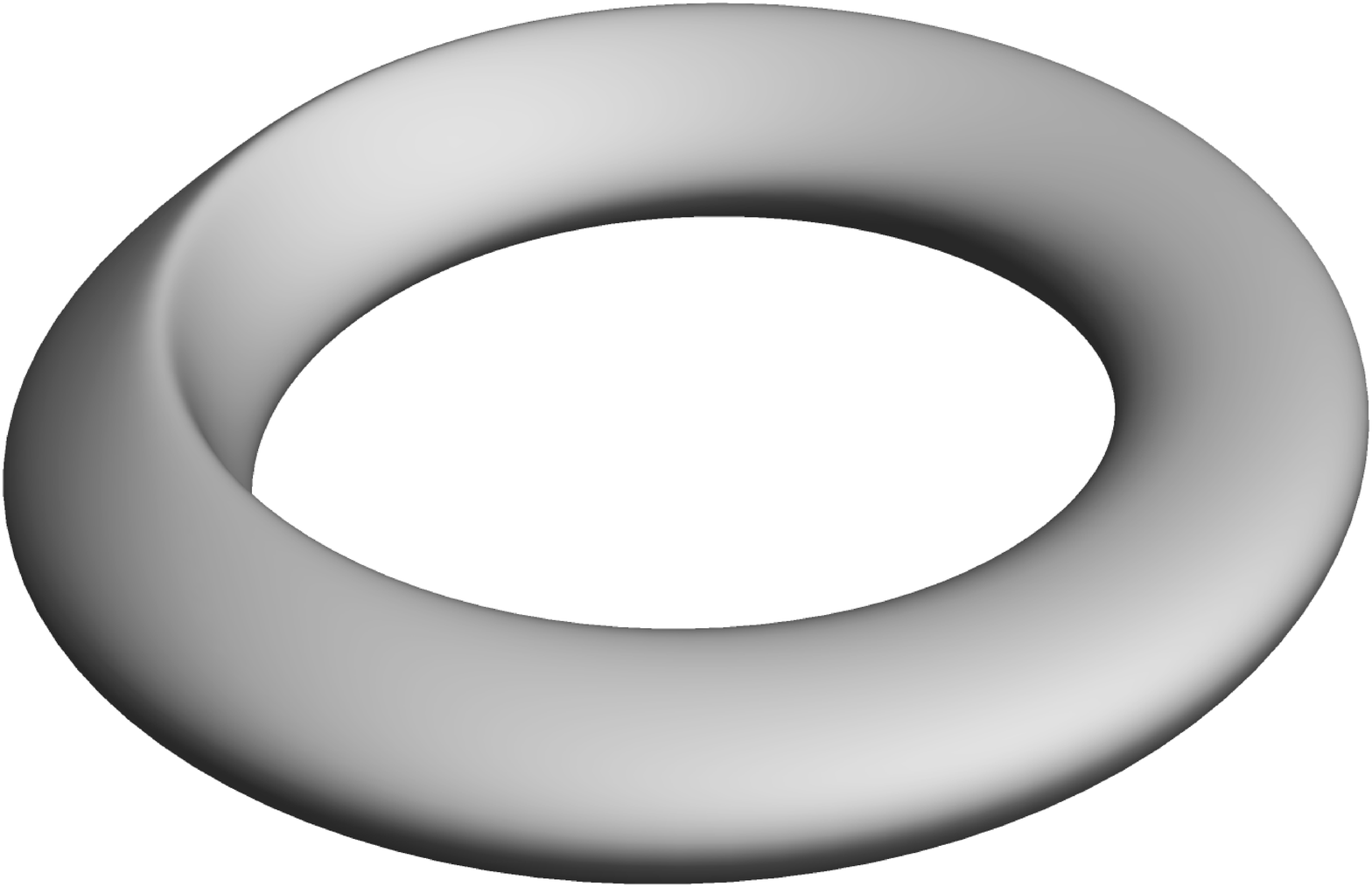}}
 \put(105,5){\includegraphics[width=10pc]{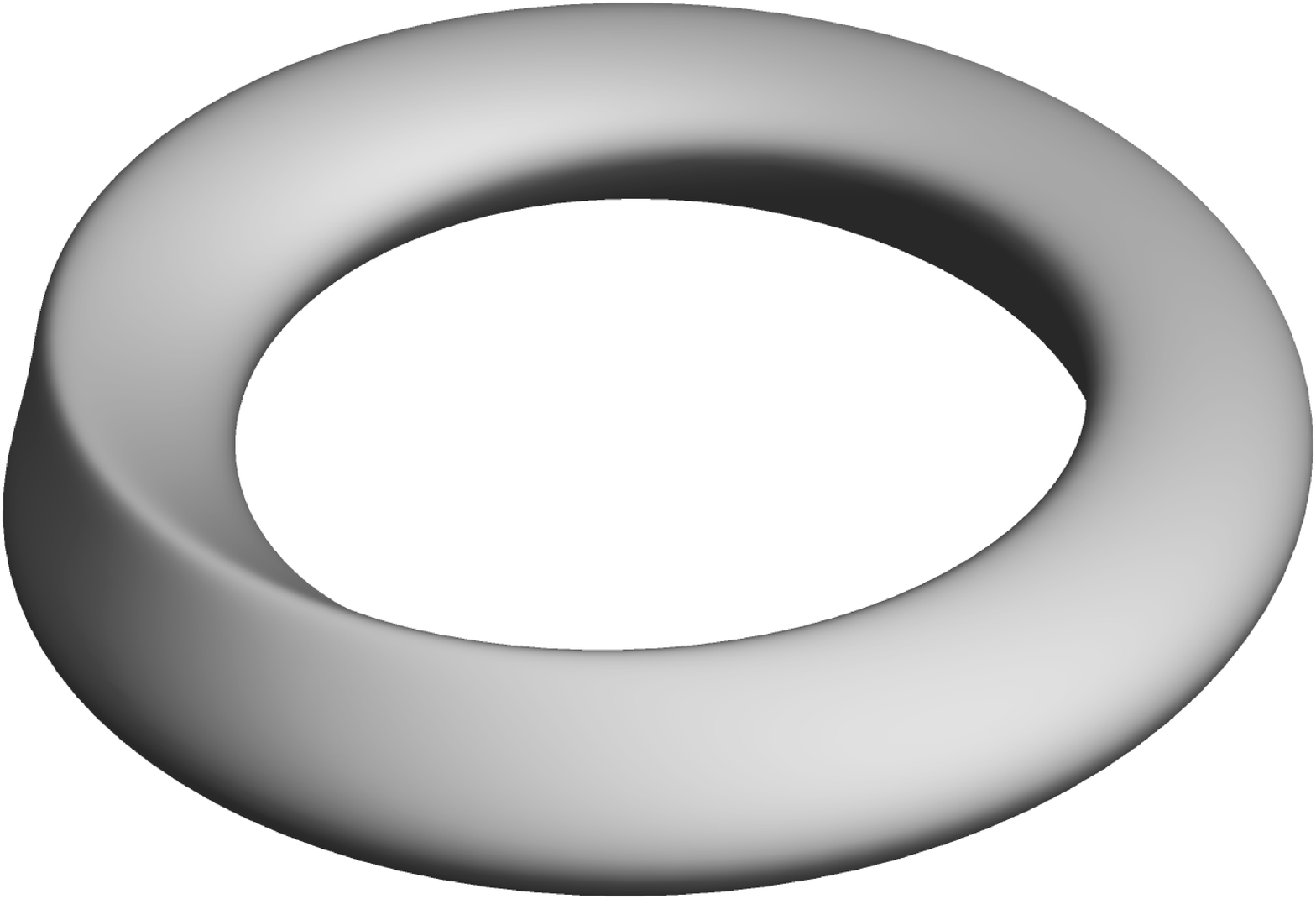}}
\fi
\put(55,95){$0\,T$}
\put(155,95){$0.25\,T$}
\put(55,10){$0.5\,T$}
\put(155,10){$0.75\,T$}
\end{picture}
\caption{\label{labelFig2}Time evolution of an appropriate surface for $T$-periodical solution with $m=1$, $n=1$.}
\end{minipage}
\end{figure}

The toroidal coordinate system $\{\kappa,\upsilon,\varphi\}$ has
the following nonzero components of diagonal metric tensor:
\begin{equation}
\label{315028981}
\metr_{\kappa\kappa} = \metr_{\upsilon\upsilon} = \dfrac{\rho_\circ^2}{(\cosh \kappa - \cos \upsilon)^2}\;,\quad
\metr_{\varphi\varphi}=\dfrac{\rho_\circ^2\,\sinh^2 \kappa}{(\cosh \kappa - \cos \upsilon)^2}
\;,
\end{equation}
where $\rho_\circ$ is a radius of the coordinate system singular ring.
The section $\varphi = 0 \cup \varphi = \pi$ for the toroidal coordinates is shown on Fig. \ref{labelFig1}.

The linear limit of the model under consideration with $\chi = 0$ gives the linear wave equation or D'Alembert one.
We can obtain the form of this equation in toroidal coordinates.
For the harmonic time dependence $\Phi=\bar{\Phi}(\kappa,\upsilon,\varphi)\,\exp({\rm i}\,\omega\,x^0)$ with the angular frequency $\omega$,  we have the following Helmholtz equation for the space-dependent
function $\bar{\Phi}$:
\begin{multline}
\label{534236125}
\left(\cos\upsilon-\cosh\kappa\right)^2\left(\sinh^2\!\kappa \left(\frac{\p^2 \bar{\Phi}}{\p \kappa^2} + \frac{\p^2 \bar{\Phi}}{\p \upsilon^2}\right)
+\frac{\p^2 \bar{\Phi}}{\p \varphi^2}\right)
{}+{} \rho_\circ^2\,\omega^2\,\sinh^2\!\kappa\;\bar{\Phi}
\\
{}+{}
\sinh\kappa\left(\cos\upsilon-\cosh\kappa\right)
\left(
\left(\cosh\kappa\,\cos\upsilon - 1 \right)\frac{\p\bar{\Phi}}{\p\kappa}
+
\sinh\kappa\,\sin\upsilon\,\frac{\p\bar{\Phi}}{\p\upsilon}
 \right) = 0
\;.
\end{multline}

It is well known that the variables are not separated completely in this equation for the general case $\omega\neq 0$.
At the present time we do not know a solution for this linear equation.

Nonlinear equation for the model under investigation in toroidal coordinates is obtained from the
action
(\ref{35135655})
with the metric  (\ref{315028981}).
The obtained equation is very complicated for the representation in the paper.
This equation can be analyzed with the help of computer mathematical programs shush that {\it Wolfram Mathematica}.

\section{Ringed lightlike soliton}
\label{tpts}

As noted above we consider the first-order twisted lightlike soliton of extremal space-time film as photon.
Also we have the higher-order twisted lightlike solitons which may represent neutrinos.

The appropriate circinate solitons with a static part can be considered as a massive charged particles with spin and magnetic moment.
The idea is that the appropriate solutions of extremal space-time film may represent leptons.

We consider the subclass of time-periodic toroidal solutions with dependence on three variables:
\begin{equation}
\label{358922321}
\Phi(x^{0},\kappa,\upsilon,\varphi)  = \tilde{\Phi}(\theta,\kappa,\upsilon)
\;,
\end{equation}
where $\theta\eqdef \varphi - \tilde{\omega}\,x^{0}$, $\tilde{\omega}$ is the angular velocity.
Also we take the condition
$\tilde{\omega}\,\rho_\circ  = 1$
such that the phase velocity of the circular wave on the singular ring equals the speed of light.

For the case under consideration the equation contains one parameter
$\varepsilon  \eqdef \pm\chi^2/\rho_{\circ}^{2}$,
where
top and bottom signs correspond to two signatures of the metric of the model (see (\ref{exatf})).

We search the solution in the following form:
\begin{equation}
\label{373761591}
\tilde{\Phi}  = \sqrt{\cosh\kappa - \cos\upsilon}\;\bar{\tilde{\Phi}}
\;,\quad
 \bar{\tilde{\Phi}} = \sum_{i=0}^{N} \bar{\tilde{\Phi}}_{i}(\theta,\upsilon)\,\kappa^{i}
 \;,
\end{equation}
where $\bar{\tilde{\Phi}}$ is represented by the partial sum of formal power series in $\kappa$.

Each iteration gives the equation for next coefficient of the power series $\bar{\tilde{\Phi}}_{i}$.
These equations with the exception of the first one are two-order ordinary linear differential equations
with respect to
variable $\theta$.
These equations do not include the derivatives of the coefficients $\bar{\tilde{\Phi}}_{i}$ with respect to variable $\upsilon$.
We can find the solution of equation for each iteration.
Thus we can build the solution
for any order of approximation $N$.

The first equation which appear in the iteration method
has the following form:
\begin{equation}
\label{419387461}
\varepsilon\,\bar{\tilde{\Phi}}_{1}\left(\frac{\p\bar{\tilde{\Phi}}_{0}}{\p\theta}\right)^2  = 0
\;.
\end{equation}

This equation is satisfied for the linear case $\varepsilon = 0$ but we take the interest in nonlinear one $\varepsilon \neq 0$.
We take $\bar{\tilde{\Phi}}_{0}$ to be constant which provides the right behavior of the solution at space infinity as charged elementary particle.
The calculations gives for this case that the two first terms of the series can be taken in the following form:
\begin{equation}
\label{585511681}
\bar{\tilde{\Phi}}  =  \frac{1}{\sqrt{2}}\,e\,\tilde{\omega} + \kappa^{n}\,\bar{\bar{\tilde{\Phi}}}_{n}(\upsilon)\,\sin\!\left(n\,\theta - m\,\upsilon \right)
+ \dots
\;,
\end{equation}
where $e$ is an electrical charge, $n$ is the number of minimal harmonic in variable $\theta\eqdef \varphi - \tilde{\omega}\,x^0$,
$m$ is the twist parameter of the appropriate circinate lightlike soliton.
The parameter $n$ defines the number of wavelengths 
on the singular ring.
A time evolution of an appropriate surface with the same dynamical symmetry that the solution
 with $m=1$ and $n=1$ is shown on Fig. \ref{labelFig2}.

We can suppose that the different number pairs $\{m,n\}$  represents the different particles.

For example, let us write the following three first terms of the solution for $n = 1$:
\begin{equation}
\label{614922831}
\bar{\tilde{\Phi}}  =
\frac{1}{\sqrt{2}}\,e\,\tilde{\omega} + \kappa\,w\,\sin^4\!\frac{\upsilon}{2}\,\sin(\theta - m\, \upsilon )
-\kappa^2\, e\,\tilde{\omega}\,
\frac{f_1(\epsilon,\upsilon)}{f_2(\epsilon,\upsilon)}
+
\OOO{\kappa^3}{\kappa\to\infty}\,,
\end{equation}
where a value of the parameter $w$ is not defined in this step of the iterative process.

Let us transform the expression (\ref{614922831}) to spherical coordinates $\{r,\vartheta,\varphi\}$:
\begin{equation}
\label{360058031}
\bar{\tilde{\Phi}} \approx \frac{e}{r}+\dots+ \frac{w\,2 \sqrt{2} \rho_\circ^6\,   \sin (\vartheta )\, \cos^4(\vartheta )}{r^6}\,\sin\theta
+\frac{ w\,4 \sqrt{2}\, \rho_\circ^7 \, \sin\vartheta\, \cos^5\vartheta }{r^7}\,\cos\theta + \dots
\;,
\end{equation}
where the dependence on the phase $\theta\eqdef\varphi - \tilde{\omega}\,x^0$ appears in the sixth and seventh order by inverse radius
only.
Thus we can talk that the time dependence is deeply embedded in the solution.
This provides the integral convergence of energy and angular momentum at infinity,
in contrast to the case of an elementary time-periodic solution for wave equation in spherical coordinates.

\section{Conclusion}
\label{concl}

We have considered the field model of extremal space-time film.
We have obtained the model equation in toroidal coordinates.
We have considered the time-periodic solution dependent on three variables. This solution contain the circular wave
with the phase velocity on the coordinate singular ring equal the speed of light. Also the solution contain a static part.

We have proposed the iterative algorithm for obtaining this solution in the form of formal power series in toroidal variable $\kappa$.
Using this algorithm we can obtain any term of the power series.

We discover that the time dependence in the toroidal solution has a deeply embedded character.
Thus the full energy and angular momentum of the solution converge at space infinity.

Thus the soliton solution under consideration can represent a massive charged wave-particle with spin.

\section*{References}
\providecommand{\newblock}{}

\end{document}